**‡ Fermilab**

# RECENT ENHANCEMENTS TO THE MARS15 CODE [*]

N.V. Mokhov[1], K.K. Gudima[2], C.C. James[1], M.A. Kostin[1], S.G. Mashnik[3], E. Ng[4],
J.-F. Ostiguy[1], I.L. Rakhno[5], A.J. Sierk[3], S.I. Striganov[1]

[1]*Fermi National Accelerator Laboratory, MS 220, Batavia, Illinois 60510-0500, USA*
[2]*Institute of Applied Physics, Academy of Sciences of Moldova, Kishinev, MD-2028, Moldova*
[3]*Los-Alamos National Laboratory, MS B283, Los-Alamos, New-Mexico 87545, USA*
[4]*University of Illinois at Chicago, 845 W. Taylor st., Chicago, IL 60607, USA*
[5]*University of Illinois at Urbana-Champaign, 1110 W. Green st., Urbana, Illinois 61801-3080, USA*

April 29, 2004

**Abstract**

The MARS code is under continuous development and has recently undergone substantial improvements that further increase its reliability and predictive power in numerous shielding, accelerator, detector and space applications. The major developments and new features of the MARS15 (2004) version described in this paper concern an extended list of elementary particles and arbitrary heavy ions and their interaction cross-sections, inclusive and exclusive nuclear event generators, module for modelling particle electromagnetic interactions, enhanced geometry and histograming options, improved MAD-MARS Beam Line Builder, enhanced Graphical-User Interface, and an MPI-based parallelization of the code.

# RECENT ENHANCEMENTS TO THE MARS15 CODE

N.V. Mokhov[1*], K.K. Gudima[2], C.C. James[1], M.A. Kostin[1], S.G. Mashnik[3], E. Ng[4], J.-F. Ostiguy[1], I.L. Rakhno[5], A.J. Sierk[3], S.I. Striganov[1]

[1]*Fermi National Accelerator Laboratory, MS 220, Batavia, Illinois 60510-0500, USA*
[2]*Institute of Applied Physics, Academy of Sciences of Moldova, Kishinev, MD-2028, Moldova*
[3]*Los-Alamos National Laboratory, MS B283, Los-Alamos, New-Mexico 87545, USA*
[4]*University of Illinois at Chicago, 845 W. Taylor st., Chicago, IL 60607, USA*
[5]*University of Illinois at Urbana-Champaign, 1110 W. Green st., Urbana, Illinois 61801-3080, USA*
*Corresponding author: phone +1-630-840-4409, fax +1-630-840-6039, e-mail: mokhov@fnal.gov

**Abstract** – The MARS code is under continuous development and has recently undergone substantial improvements that further increase its reliability and predictive power in numerous shielding, accelerator, detector and space applications. The major developments and new features of the MARS15 (2004) version described in this paper concern an extended list of elementary particles and arbitrary heavy ions and their interaction cross-sections, inclusive and exclusive nuclear event generators, module for modelling particle electromagnetic interactions, enhanced geometry and histograming options, improved MAD-MARS Beam Line Builder, enhanced Graphical-User Interface, and an MPI-based parallelization of the code.

## INTRODUCTION

The MARS code system, developed over 30 years, is a set of Monte Carlo programs for detailed simulation of hadronic and electromagnetic cascades in an arbitrary geometry of shielding, accelerator, detector and spacecraft components with particle energy ranging from a fraction of an electronvolt up to 100 TeV[1]. The growing needs of new accelerator and space projects with their respective experiments, stimulated new developments geared towards enhanced modelling of elementary particle and heavy ion interactions during transport in realistic micro and macro systems. Here, the most recent developments to the new MARS15 (2004) version, further increasing the code reliability, applicability and user friendliness, are highlighted.

## EVENT GENERATORS

A set of phenomenological models, used as a default for inclusive particle production in inelastic nuclear interactions at energies above 5 GeV, was further refined and improved for leading baryons, high transverse momentum pions and neutral kaons. A new model for fast nucleon production deals with four kinematic regions: (i) resonance region, where target nucleon is excited to a baryon resonance; (ii) diffraction dissociation with a target nucleon excited to a high-mass state; (iii) fragmentation where a leading nucleon exhibits an exponential dependence on transverse momentum; and (iv) central region with a Gaussian behavior of transverse momentum. A dependence on transverse momentum in a nuclear modification factor $R$ $(pA/pp)$ was added. Fig. 1 shows an example of a new model prediction against data[2] for 100-GeV $pA$ collisions on different nuclei.

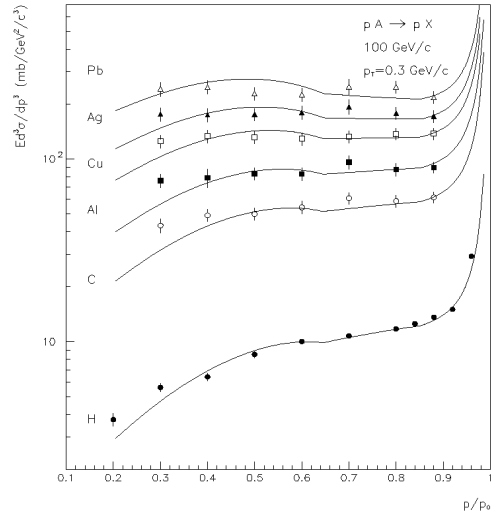

Figure 1. The invariant cross section for $pA \rightarrow pX$ reactions at 100 GeV/c in MARS15 (lines) *vs* data[2].

The 2003 version of the improved Cascade-Exciton Model[3] code, CEM03, combined with the Fermi break-up model, the coalescence model, and an improved version of the Generalized Evaporation-fission Model (GEM2) is used as a default for hadron-nucleus interactions below 5 GeV. The 2003 version of the Los Alamos Quark-Gluon String Model[4] code, LAQGSM03, was implemented into MARS15 for particle and heavy-ion projectiles at 10 MeV/A to 800 GeV/A. Details of these two event generators can be found in Ref. (5). This implementation provides the capability of full theoretically consistent modelling of exclusive distributions of secondary particles, spallation, fission and fragmentation products. Fig. 2 shows a mass distribution of selected nuclides in *pU* interactions at 1 GeV, while Fig. 3 shows inclusive π⁻ spectra in *Au+Au* collisions at 4 GeV/A.

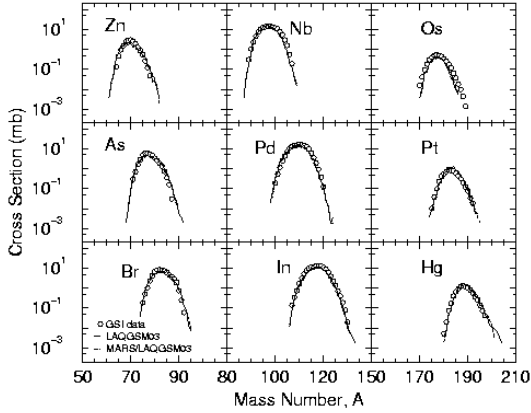

Figure 2. Nuclide mass yield in *p+U* interaction at 1 GeV as calculated with original LAQGSM03 and that implemented into MARS15, and measured in Ref. (6).

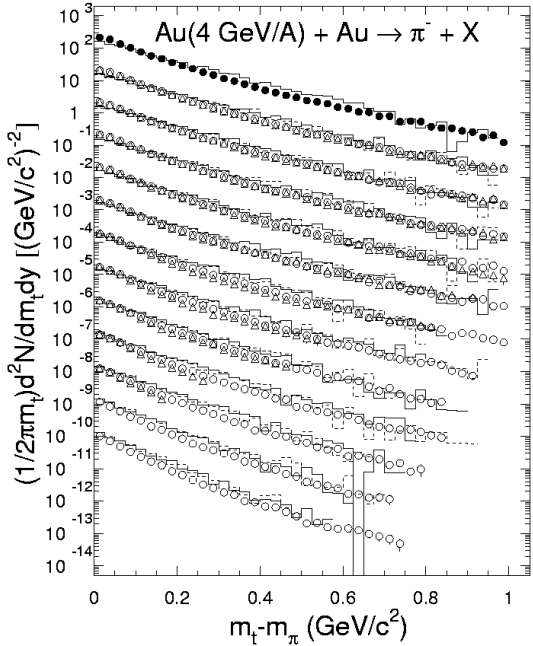

Figure 3. Invariant π⁻ yield per central *Au+Au* collision at 4 GeV/A as calculated with LAQGSM03 (histograms) and measured in Ref. (7) (symbols). Solid lines and open circles is forward production, dashed lines and open triangles is backward production. Midrapidity (upper set) is shown unscaled, while the 0.1 unit rapidity slices are scaled down by successive factors of 10.

## PARTICLES AND CROSS SECTIONS

Neutral kaons, antineutron and hyperons (Table I) are now transported in the MARS15 code, all with their corresponding decay modes. Arbitrary heavy ions with atomic mass $A$ and charge $Z$ are now fully treated by MARS15[8]. Their *ID* are coded as *ID=1000Z+A-Z*.

Table I. Elementary particles transported in MARS15.

| Nucleons & Mesons | | Gauge & Leptons | | Hyperons | |
|---|---|---|---|---|---|
| $p$ | $\bar{p}$ | $\gamma$ | | $\Lambda$ | $\bar{\Lambda}$ |
| $n$ | $\bar{n}$ | $e^+$ | $e^-$ | $\Sigma^+$ | $\bar{\Sigma}^+$ |
| $\pi^+$ | $\pi^-$ | $\mu^+$ | $\mu^-$ | $\Sigma^0$ | $\bar{\Sigma}^0$ |
| $\pi^0$ | | $\nu_e$ | $\bar{\nu}_e$ | $\Sigma^-$ | $\bar{\Sigma}^-$ |
| $K^+$ | $K^-$ | $\nu_\mu$ | $\bar{\nu}_\mu$ | $\Xi^0$ | $\bar{\Xi}^0$ |
| $K^0$ | $\bar{K}^0$ | | | $\Xi^-$ | $\bar{\Xi}^-$ |
| $K^0_L$ | $K^0_S$ | | | $\Omega^-$ | $\bar{\Omega}^-$ |

As in the previous version, total and elastic cross sections of hadron-nucleon interactions for ordinary hadrons are described using corresponding fits to experimental data[1]. Cross sections for hyperon-nucleon interactions are described via the ordinary hadron cross sections using the additive quark model rules. At energies above 5 GeV, such an approach agrees well with data. At lower energies, the hyperon-nucleon cross sections are very close to proton-nucleon ones. Hadron-nucleus total and inelastic cross sections at energies above 5 GeV are calculated in the code using the Glauber model. At lower energies, parametrizations to experimental data are used. This description agrees well with data. For neutral kaons, cross sections on both nucleon and nucleus targets are calculated using the relation based on isospin and hypercharge conjugation: $\sigma^{K_{S,L}} = (\sigma^{K^+} + \sigma^{K^-})/2$. Total and inelastic cross sections for heavy ion nuclear interactions are based on the JINR algorithm[9] (see analysis in Ref. (8)).

Table 2 gives a comparison to experiment[10] of the calculated total neutron yield below 14.5 MeV from a lead cylinder of 20-cm diameter and 60-cm thick irradiated by 0.5 to 3.65 GeV/A light ion beams. As with other projectiles and energies described in Ref. (8) agreement is quite good.

Table 2. Neutron yield from a lead cylinder for deuteron and ¹²C beams calculated with MARS15 (C) vs data[10] (E).

| E (GeV/A) | | 0.52 | 0.99 | 1.28 | 1.88 | 3.65 |
|---|---|---|---|---|---|---|
| d | C | 28.3 | 50.8 | 62.6 | 85.3 | 144 |
| | E | 24.4 | 52.1 | 56.4 | 89.4 | 156 |
| ¹²C | C | 64.1 | 162 | 219 | 326 | 601 |
| | E | - | 132 | 198 | 306 | 630 |

## ELECTROMAGNETIC PROCESSES

A new algorithm[11] for modelling correlated ionization energy loss and multiple Coulomb scattering was implemented for arbitrary mixtures. It takes into account arbitrary projectile and nucleus charge distributions, exact kinematics of projectile-electron interactions and accurately treats both soft and hard collisions. Calculated correlations between energy loss and scattering are quite substantial for low-Z targets. The mean stopping power used for normalization of ionization energy loss through a step is calculated using a Bethe-Bloch formalism with corresponding low- and high-energy corrections implemented previously[1,12]. Two additional corrections specific for heavy ions have been implemented into the code to describe better the Barkas effect and take into account electron capture by low-energy ions, providing a few percent accuracy. A comparison of new MARS15 results with our modification BMS[13] of the Moliere theory - which takes into account nuclear screening and projectile-electron interactions - is shown in Fig. 4 along with the Rossi and Moliere distributions (see Ref. (13)) for a 50 GeV/c proton on a 10 g/cm² thick uranium target. The figure shows importance of the BMS modifications to the theory. Some difference at very large angles between the BMS model and new algorithm is due to the fact that the former uses a simplified Gaussian form-factor while the later a more precise Fermi distribution of charge in the target nucleus. Radiative processes for single-charged particles and heavy ions – bremsstrahlung and direct pair production – are modelled directly[1].

## GEOMETRY, HISTOGRAMING AND TAGGING

In addition to the widely used *standard*, *non-standard* and *extended* geometries, two newly developed geometry options in MARS15 enhance the code's capabilities and make it compatible – in a geometry sense – with two other simulation communities, MCNP and FLUKA: MARS15 can now read in an input geometry description in the MCNP and FLUKA formats. The *extended* geometry sector was drastically improved. *Extended* zones are constructed from a set of contiguous or overlapping geometrical shapes, currently, boxes, spheres, cylinders, truncated cones and tetrahedra. A variety of new features, such as up to $10^5$ extended volumes, subdivision of volumes into sub-regions, up to 500 arbitrary transformation matrices etc, are present. Automatic adjustments of step sizes along a particle track prevents small regions within a large volume from being skipped over. All five geometry options can co-exist in a setup description. Volumes of all regions in MARS15 are auto-calculated using a short session of the program. A corresponding output file provides calculated volumes with statistical errors, and is directly linked to the main code. Fig. 5 shows as an example of geometry description, the central part of the CMS detector (see next section on GUI).

A new user-friendly flexible XYZ-histograming module allows scoring numerous distributions – total and partial particle fluxes, star density, prompt and residual dose rates, particle spectra etc. – in boxes arbitrary positioned in a 3D system, independently of the geometry description.

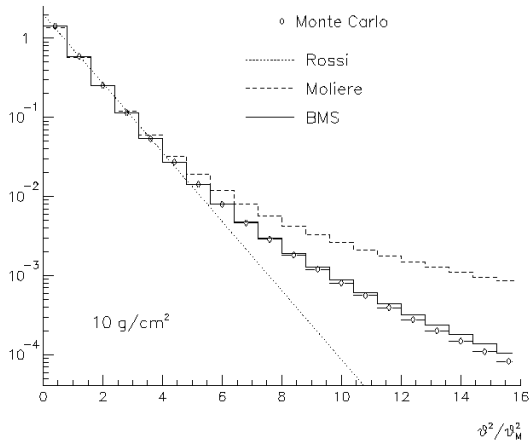

Figure 4. Angular distribution of 50 GeV/c protons after a uranium absorber.

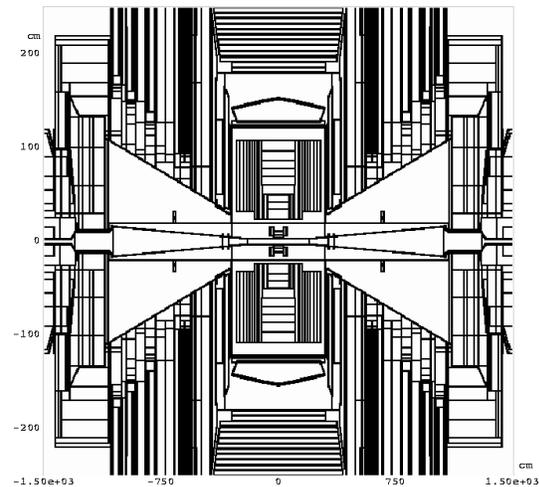

Figure 5. A fragment of the FLUKA-MARS15 model of the CMS detector as seen in MARS-GUI-SLICE.

A refined tagging module in MARS15 allows one to tag the origin of a given signal/tally: geometry, process and phase-space. A list of the built-in materials has been extended up to 153 with a separate treatment of gaseous and liquid states of some elements.

## GRAPHICAL USER INTERFACE

The GUI is invaluable for debugging geometry, materials and magnetic field descriptions, simulated processes and calculated results. The existing Tcl/Tk-based 2D MARS-GUI-SLICE functionality was further improved and extended to 3D[14], which further extends the power of visualization of the modelled system. It is based on the Open Inventor graphics library, integrated with MARS-GUI-SLICE. It has also been re-implemented using the C++ based Qt-toolkit. The new 3D display capability was developed using open source libraries that should allow redistribution of the code and/or binaries to the MARS user community, free of commercial licensing terms.

## BEAM LINE BUILDER

The MAD-MARS Beam Line Builder (MMBLB) builds beam line and accelerator models in the MARS15 format by using a beam line description of the MAD accelerator design code. It has been substantially extended[15]: (i) the set of supported element types includes now almost all the elements supported by MAD; (ii) an arbitrary number of beam lines - arbitrary positioned and oriented - can be put in a MARS15 model; (iii) more sophisticated algorithms and new data structures enable more efficient searches through the beam line geometry; (iv) tunnel geometry can now follow the beam line or be described independently of it.

## MPI PARALLELIZATION

A parallel processing option has been developed and implemented into the MARS15 code[16]. It is based on the Message Passing Interface (MPI) libraries. Parallelization is job-based, i.e. the processes, replicating the same geometry of the setup studied, run independently with different initial seeds. A unique master process – also running event histories – collects intermediate results from an arbitrary number of slaves and calculates the final results when a required total number of events has been processed. Intermediate results are sent to the master on its request generated in accordance with a scheduling mechanism. The algorithm was tested on Unix and Linux clusters and its performance is described in Ref. (16).

This work was supported by the Universities Research Association, Inc., under contract DE-AC02-76CH03000 with the U.S. Department of Energy, and in part by the Moldovan-U.S. Bilateral Grants Program, CRDF Projects MP2-3025 and MP2-3045, the NASA ATP01 Grant NRA-01-01-ATP-066, and the Higher Education Cooperative Act Grant of the Illinois Board of Higher Education.

## REFERENCES

1. Mokhov, N. V. *The MARS code system user's guide.* Fermilab-FN-628, Fermi National Accelerator Laboratory (1995); Mokhov, N. V., Krivosheev, O. E. *MARS code status.* Proc. Monte Carlo 2000 Conf., p. 943, Lisbon, October 23-26, 2000, Fermilab-Conf-00/181 (2000); Mokhov, N. V. *Status of MARS code.* Fermilab-Conf-03/053, Fermi National Accelerator Laboratory (2003); http://www-ap.fnal.gov/MARS/.

2. Barton, D.S., et al. (34 authors). *Experimental study of the A dependence of inclusive hadron fragmentation.* Phys. Rev. **D27** (11), 2580-2599 (1983).

3. Mashnik, S. G., and Sierk, A. J. *CEM2k—recent developments in CEM.* Proc. Conf. on Accelerator Applications, Washington DC, USA, pp. 328-341 (2000), E-print: nucl-th/0011064.

4. Gudima, K. K., Mashnik, S. G., and Sierk, A. J. *User manual for the code LAQGSM,* LA-UR-01-6804, Los Alamos National Laboratory (2001).

5. Mashnik, S.G., Gudima, K.K., Prael, R.E., Sierk, A.J. *Analysis of the GSI A+p and A+A spallation, fission, and fragmentation measurements with the LANL CEM2k and LAQGSM codes.* Proc. Workshop on Nuclear Data for the Transmutation of Nuclear Waste, GSI, Darmstad, September 1-5, 2003, Los Alamos LA-UR-04-1873 (2004), E-print: nucl-th/0404018; Mashnik, S.G., Gudima, K.K., Sierk, A.J, Prael, R.E. *Improved intranuclear cascade models for the codes CEM2k and LAQGSM,* Los Alamos LA-UR-04-0039 (2004); Mashnik, S.G., Gudima, K.K., Prael, R.E., Sierk, A.J. *Recent developments in LAQGSM.* In these proceedings.

6. Taieb, J., et al. (18 authors). *Measurement of nuclide cross-sections of spallation residues in 1 A GeV ²³⁸U +proton collisions.* Nucl. Phys. **A724**, 413 (2003); Bernas, M. et al. (19 authors). *Fission-residues produced in the spallation reaction 238U+p at 1 A GeV.* Nucl. Phys. **A725**, pp. 213-253 (2003).

7. Klay, J.L., et al. (53 authors). *Charged pion production in 2 to 8 A GeV central Au+Au collisions.* Phys. Rev. **C68**, 054905 (2003).

8. Mokhov, N.V., Gudima, K.K., Mashnik, S.G., Rakhno, I.L., Striganov, S.I. *Towards a heavy-ion transport capability in the MARS15 code.* In these proceedings.

9. Barashenkov, V. S., Kumawat, H. *Integral nucleus-nucleus cross-sections.* JINR E2-2003-128, Joint Inst. for Nuclear Research (2003).

10. Vassil'kov, R.G., Yurevich, V.I. *Neutron emission from an extended lead target under the action of light ions in GeV region.* Proc. ICANS-XI Conf, KEK, Tsukuba, Japan, October 22-26, 1990, vol. 1, p. 340.

11. Striganov, S.I. *On the theory and simulation of multiple Coulomb scattering of heavy particles.* In these proceedings.

12. Groom, D. E., Mokhov, N. V., and Striganov, S. I. *Muon stopping power and range tables 10 MeV–100 TeV.* Atomic Data and Nucl. Data Tables. **78** (2), 183-356 (2001).

13. Baishev, I.S., Mokhov, N.V., Striganov, S.I. *On the effect of the finite size of the nucleus in the theory of multiple Coulomb scattering.* Sov. J. Nucl. Phys. **42** (5), 745-749 (1985).

14. Rzepecki, J.P., Kostin, M.A., Mokhov, N.V. *3D visualization for the MARS14 code.* Fermilab-TM-2197, Fermi National Accelerator Laboratory (2003).

15. Kostin, M.A., Krivosheev, O.E., Mokhov, N.V., Tropin, I.S. *An improved MAD-MARS Beam Line Builder: User's guide.* Fermilab-FN-738, Fermi National Accelerator Laboratory (2003).

16. Kostin, M.A., Mokhov, N.V. *Parallelizing the MARS15 code with MPI for shielding applications.* In these proceedings.